\documentclass[vecphys]{svmult}
\usepackage{epsfig}
\usepackage{amsmath,amssymb}
\usepackage{graphicx}

\title{Wigner function quantum molecular dynamics}
\author{Vladimir Filinov\inst{1,2} \and Michael Bonitz\inst{2} \and Alexei Filinov\inst{2} \and Volodymyr Golubnychiy\inst{2}}

\authorrunning{V. Filinov \and M. Bonitz \and A. Filinov \and V. Goluybnychiy }
\institute{Institute for High-Energy Density, Russian Academy of Sciences,
Moscow 127412, Russia
\and
Institut f\"ur Theoretische Physik und Astrophysik, Christian-Albrechts-Universtit\"at, 24098 Kiel
}

\begin{document}

\maketitle

\section{Introduction}

Classical molecular dynamics (MD) is a well established and powerful
tool in various fields of science, e.g. chemistry, plasma physics, cluster
physics and condensed matter physics. Objects of investigation are few-body
systems and many-body systems as well. The
broadness and level of sophistication of this technique is
documented in many monographs and reviews, see for
example~\cite{Allan,Frenkel,mdhere}. Here we discuss the extension of MD to quantum systems (QMD).
There have been many attempts in this direction which differ from one another, depending on the type of
system under consideration. One direction of QMD has been developed for condensed matter systems
and will not discussed here, e.g. \cite{fermid}. In this chapter we are dealing with unbound electrons as they occur
in gases, fluids or plasmas. Here, one strategy is to replace classical point particles by wave packets, e.g.
\cite{fermid,KTR94,zwicknagel06} which is quite successful. At the same time, this method struggles with problems related to
the dispersion of such a packet and difficulties to properly describe strong electron-ion interaction and
bound state formation. We, therefore, avoid such restrictions and consider a completely general alternative
approach. We start discussion of quantum dynamics from a general consideration of quantum
distribution functions.

\section{Quantum distribution functions}\label{wf_s}
There exists a variety of different representations of quantum
mechanics including the so-called
{\em Wigner representation} which involves a class of functions
depending on coordinates and momenta. In the classical
limit, the Wigner distribution $F^W$ turns into the phase space distribution $f$ known
from classical statistical mechanics. In contrast to $f$, the Wigner function
may be non-positive which is a consequence of the coordinate-momentum (Heisenberg) uncertainty.
This will lead to a modification of particle trajectories which we discuss below in Sec.~\ref{qd_s}.
An important property of the distribution functions is that
they can be used to compute the expectation value of an arbitrary physical
observable, $\langle A \rangle$, defined by the operator $\hat A(\hat p, \hat
q)$~\cite{QKTB},
\begin{eqnarray}
\langle A \rangle (t)= \int dp dq \, A^W(p,q) F^W(p,q,t), \quad
1 = \int dp dq F^W(q,p,t),
\label{mA_f}
\end{eqnarray}
where $A^W(p,q)$ is a scalar function and, for simplicity, we consider the $1$D case
(generalization to higher dimensions and $N$ particles is
straightforward by re-defining the coordinate and momentum as vectors, $q=\{\mathbf q_1,\ldots,\mathbf q_N\}$,
$p=\{\mathbf p_1,\ldots,\mathbf p_N\}$). $F^W$ is defined via the nonequilibrium N-particle density operator $\hat \rho$
in coordinate representation (i.e. the density matrix),
\begin{eqnarray}
F^W(p,q,t)=\frac{1}{2\pi \hbar} \int d\nu  \, \left\langle q
+\frac{\nu}{2}|\hat \rho| q -\frac{\nu}{2} \right\rangle \, e^{-i\nu p}
\label{Fw_coord},
\end{eqnarray}
and $A^W(p,q)$ is analogously defined from the coordinate represention of $\hat A$.

We now consider the time evolution of the WF under the influence of a general hamiltonian of the form
\begin{eqnarray}
\hat H=\sum_{j=1}^N \frac{\hat p_i^2}{2m} +  \sum_{i=1}^N \tilde
V(q_i)+ \sum_{i<j} V(q_i,q_j),
\label{h}
\end{eqnarray}
where $\tilde V(q_i)$ and $V(q_i,q_j)$ denote an external and an interaction potential, respectively.
The equation of motion of $F^W$ has the form~\cite{Wig,QKTB}, some explanations are given in Sec.~\ref{qd_s},
\begin{equation}
\frac{\partial F^W}{\partial t}+ \frac{{\vec p}}{m} \cdot {\vec\nabla}_q F^W =
\int_{-\infty }^{\infty }ds\, F^W\left(
p-s,q,t\right) \, \tilde{\omega} \left( s,q,t\right),
\label{qwl}
\end{equation}
where the function
\begin{eqnarray}
\tilde{\omega} \left( s,q,t\right) =\frac{2}{\pi \hbar ^{2}}\int
dq^{\prime }V\left( q-q^{\prime },t\right) \sin \left(
\frac{2sq^{\prime }}{\hbar } \right)
\end{eqnarray}
takes into account the non-local contribution of the potential
energy in the quantum case. Equivalently, expanding the integral around $q'=0$, Eq.~(\ref{qwl})
can be rewritten with an infinite sum of local potential terms
\begin{equation}
\frac{\partial F^W}{\partial t}+\frac p m \frac{\partial F^W}
{\partial q} = \sum_{n=0}^\infty \frac{(\hbar/2i)^{2n}}{(2n+1)!}
\left(\frac{\partial^{2n+1}V}{\partial q^{2n+1}},
\frac{\partial^{2n+1}F^W}{\partial p^{2n+1}}\right),
\label{WigEvol3}
\end{equation}
where $\left(\ldots,\ldots\right)$ denotes
the scalar product of two vectors which for an $N$-particle system contain
$3N$ components.

If the potential does not contain terms higher than the second
power of $q$, i.e. $\frac{\partial^n V}{\partial q^n}|_{n\geq 3}= 0$,
then Eq.~(\ref{WigEvol3}) is simplified and reduces to the classical Liouville equation
for the distribution function $f$, i.e.
\begin{equation}
\frac{\partial f}{\partial t}+\frac p m \frac{\partial f}
{\partial q} = \frac{\partial V}{\partial q}\frac{\partial f}
{\partial p}.
\label{ClasDistribEq}
\end{equation}
The Wigner function must satisfy a number of conditions
\cite{Tatarsky1983}, therefore, the initial function $F^W(q,p,0)$
cannot be chosen arbitrarily. Even if $F^W(q,p,t)$ satisfies the
classical equation~(\ref{ClasDistribEq})
it nevertheless describes the evolution of a quantum distribution because a
properly chosen initial function $F^W(q,p,0)$ contains, in general, all
powers of $\hbar$. In particular, the
uncertainty principle holds for average values of operators calculated with
$F^W(q,p,0)$ and $F^W(q,p,t)$.

One can rewrite Eq.~(\ref{WigEvol3}) in the form analogous to the classical
equation~(\ref{ClasDistribEq}) by replacing $V$
by a new effective potential $V_{\rm eff}$ defined as
\begin{eqnarray}
\frac{\partial V_{\rm eff}}{\partial q}\frac{\partial F^W}{\partial p}
= \frac{\partial V}{\partial q}\frac{\partial F^W}{\partial p}-
\frac{\hbar^2}{24}\frac{\partial^3 V}{\partial
q^3}\frac{\partial^3 F^W} {\partial p^3}+\cdots
\label{DefVeff}
\end{eqnarray}
The classical Liouville equation
~(\ref{ClasDistribEq}) can be efficiently solved with the {\em method of
characteristics}, see e.g. Ref.~\cite{ourbook}, and this is also
the basis of our QMD approach where an ensemble of
``classical'' (Wigner) trajectories is used to solve (numerically)
the quantum Wigner-Liouville equation (\ref{qwl}) which will be discussed
in Sec.~\ref{qd_s}. The time-dependence of the trajectories is
given by ``classical'' equations of motion
\begin{equation}
\frac{\partial q}{\partial t}=\frac{p}{m}, \quad \frac{\partial
p}{\partial t}=-\frac{\partial V_{\rm eff}(p,q,t)}{\partial q},
\label{eff_eq}
\end{equation}
Of course, a direct solution of Eq.~(\ref{eff_eq}) with the definition
Eq.~(\ref{DefVeff}) is only useful if the series is rapidly converging and there is only a
small number of non-zero terms.

However, there is also a
principle difficulty with this approach, if the series of terms
with the potential derivatives is not converging which is the case, e.g.,
for a Coulomb potential (at zero distance). There are at least three
solutions. The first is to solve the Wigner-Liouville equation by
using Monte Carlo techniques~\cite{filmd0}$^-$ \cite{filmd2},
which is discussed below in Section \ref{qd_s}.
The second is to replace the original potential on the r.h.s. of Eq.~(\ref{DefVeff})
by some model potential which has a finite number of nonzero
derivatives as it is done e.g. in Ref.~\cite{Arkhipov}.
The third approach is to perform a suitable average of $V_{\rm eff}$, e.g. over a thermal ensemble
of particles. This has been done both for external potentials and also for two particle interaction.
This latter case of an effective quantum pair potential and its use in classical MD is disscused in the next section.

\section{Semiclassical MD simulation}\label{MDsimulation}

{\em Quantum pair potentials.} In order to obtain an effective pair potential which is
regularized at zero by quantum effects
 (finite at zero interparticle distance), we consider Eq.~(\ref{qwl}) for 2
 particles. Assuming further thermodynamic equilibrium (with a given temperature $k_BT=1/\beta$), spatial homogeneity and
 neglecting 3-particle correlations, one can solve for the two-particle Wigner function
$F^W_{12}= F^{EQ}_{12}(r_1,p_1,r_2,p_2,\beta)\approx F^{EQ}_{12}(r_1-r_2,p_1,p_2,\beta)$.
This is now re-written as in the canonical case \cite{QKTB},
$F^{EQ}_{12}(r_1-r_2,p_1,p_2,\beta) \equiv F^{EQ}_{1}(p_1,\beta)F^{EQ}_{2}(p_2,\beta)
e^{-\beta V^Q_{12}}$,
which defines the desired quantum pair potential $V^Q_{12}$.

The first solution for $V^Q_{12}$ was found by Kelbg in the  limit of weak coupling \cite{Ke63}
and has the form of Eq.~(\ref{impr_kelbg}) with $\gamma_{ij}\rightarrow 1$, for details and
references, cf. \cite{ourbook,fil_prb}. The Kelbg potential (or slightly modified versions) is
widely used in numerical simulations of dense plasmas, e.g.~\cite{HansenMcdoald81,filinov2003,KTR94,vova01}.
It is finite at zero distance correctly capturing basic quantum diffraction effects which prevent
any divergence. However, the absolute value at $r=0$ is incorrect which has lead to the derivation
of further improved potentials, e.g. \cite{wagenknecht01,ourbook,fil_prb} and references therein.
Here we use the {\em improved Kelbg potential} (IKP),
\begin{eqnarray}
\Phi\left({r}_{{ij}},\beta\right) =\frac{q_{i}q_{j}}{r_{ij}}\,
\left[1-e^{-\frac{r_{ij}^{2}}{\lambda_{ij}^{2}}}+
\sqrt{\pi}\frac{r_{ij}}{\lambda_{ij}\gamma_{ij}}
\left(1-{\textrm{erf}}\left[\gamma_{ij}\frac{r_{ij}}{\lambda_{ij}}\right]\right)\right],
\label{impr_kelbg}
\end{eqnarray}
where $x_{ij}=|{\textbf{r}}_{ij}|/\lambda_{ij}$,  $\lambda_{ij}^{2}=\frac{\hbar^{2}\beta}{2\mu_{ij}}$ and
$\mu_{ij}^{-1}=m_{i}^{-1}+m_{j}^{-1}$, which contains an additional
free parameter  $\gamma_{ij}$ which can obtained from a fit to the exact solution of the two-particle
problem \cite{fil_prb}.

\noindent {\em MD Simulations.} We have performed extensive MD simulations of dense partially
ionized hydrogen in thermodynamic equilibrium using different IKP for electrons with different spin 
projections. To properly account for the long-range character of the
potentials, we used periodic boundary conditions with the standard Ewald procedure \cite{mdhere}.
The number of electrons and protons was $N=200$.  Our MD simulations use standard Runge-Kutte or Verlet algorithms 
\cite{mdhere} to solve Newton's equations Eq.~(\ref{eff_eq}), where $V_{\rm eff}$ is replaced by the IKP.
Because of the temperature dependence of the IKP we applied a
temperature scaling at every time step for all components
separately (for protons and two sorts of electrons) to guarantee
a constant temperature of all components in our equilibrium simulations.
In each simulation the system was equilibrated for at least
$10^{4}$ MD steps, only after this the observables have been computed.

\begin{figure}[htb]
\includegraphics[%
  scale=0.28, angle=0]{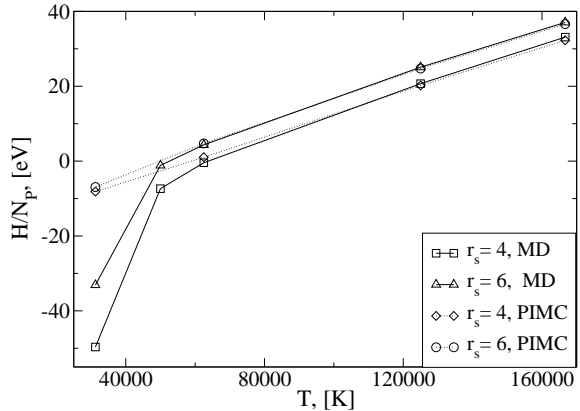}
\hspace{0.3cm}
\begin{minipage}[t]{.3\textwidth}
\vspace{-4cm}
\caption{Internal energy per hydrogen atom at $r_{s}=4$ and $r_{s}=6$ versus temperature, 
MD results are compared to restricted PIMC simulations \cite{fil_prb,golub2}.}
\label{fig:Et}
\end{minipage}
\end{figure}

In Fig.~\ref{fig:Et} we plot the internal
energy per atom as a function of temperature for two densities and
compare it to path integral Monte Carlo (PIMC) results \cite{fil_prb,golub2}.
The density is given by the Bruckner parameter $r_{s}=\bar{r}/a_B$,
where $\bar{r}$ is average interparticle distance and $a_B$ denotes the Bohr radius.
For high temperatures and weak coupling, $\Gamma=e^2/(\bar{r}k_BT) < 1$ 
(fully ionized plasma), the two simulations coincide
within the limits of statistical errors. If we use the original Kelbg potential, at 
temperatures below 300,000 K (approximately 2 times the binding energy), the MD results 
start to strongly deviate from the PIMC results. In contrast the IKP fully agrees the PIMC data even 
at temperatures far below the hydrogen binding energy $Ry$, where the plasma is dominated by atoms,
which is a remarkable extension of `semiclassical' MD into the theoretically very difficult regime 
of moderate coupling, moderate degeneracy and partial ionization.

Interestingly, even bound states can be analyzed in our simulations by folowing
the electron trajectories. At $T < 1 Ry$, we
observe an increasing number of electrons undergoing strong
deflection (large angle scattering) on protons and eventually
performing quasi-bound trajectories. Most of these electrons
remain ``bound'' only for a few classical orbits and then leave
the proton again. Averaged over a long time, our simulations are
able to reveal the degree of ionization of the plasma. 
For temperatures below approximately $50,000K$ (which is close to the
binding energy of hydrogen molecules), the simulations cannot be applied. 
Although we clearly observe molecule formation (see below), there also appear 
clusters of several molecules which is unphysical in the present conditions and is caused 
by the approximate (two-particle) treatment of quantum effects in the IKP. This turns out 
to be the reason for the too low energy in Fig.~\ref{fig:Et} at low temperatures. 

\begin{figure}[htb]
\includegraphics[%
  scale=0.29, angle=-90]{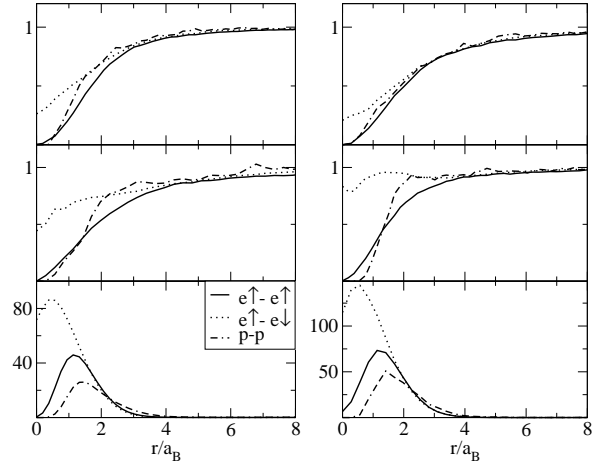}
\hspace{.2cm}
\begin{minipage}[t]{.25\textwidth}
\caption{Electron-electron and proton-proton pair
distribution functions for a correlated hydrogen plasma with
$r_{s}=4$ (left row) and $r_{s}=6$ (right row) for $T=125,000K,
61,250K$, and $31,250K$ (from top to bottom).}
\label{fig:Ee_pp_pdf}
\end{minipage}
\end{figure}

\begin{figure}[htb]
\includegraphics[%
  scale=0.27, angle=-90]{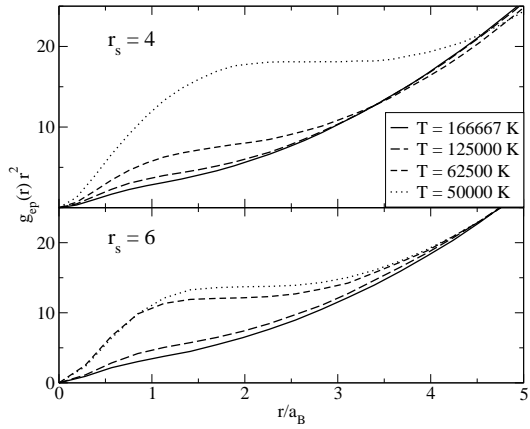}
\hspace{0.3cm}
\begin{minipage}[t]{.3\textwidth}
\caption{Electron-proton pair distribution functions multiplied
by $r^{2}$ as function of e-p distance at $r_s$= 4 (top) and
$r_s$ = 6 (bottom) and four temperatures.}
\label{fig:Ep_pdf}
\end{minipage}
\end{figure}

Let us now turn to a more detailed analysis of the spatial
configuration of the particles. In Fig.~\ref{fig:Ee_pp_pdf} the pair 
distribution functions of all particle species with
the same charge are plotted at two densities. Consider first the
case of $T=125,000$ K (upper panels). For both densities, all
functions agree qualitatively, showing a depletion at zero
distance due to Coulomb repulsion. Besides, there are differences
which arise from the spin properties. Electrons with the same spin
show a ``Coulomb hole'' around $r=0$ which is broader than
the one of the protons due to the Pauli principle (additional repulsion 
of electrons with the same spin projection).
This trend is reversed at low temperatures, see the middle panel,
which is due to the formation of hydrogen atoms and molecules. In
this case, electrons (i.e., their classical trajectories) are ``spread out''
around the protons giving rise to an increased probability of
close encounters of two electrons belonging to different atoms
compared to two protons.

Now, compare electrons with parallel vs. electrons with
anti-parallel spins. In all cases, we observe a significantly
increased probability to find two electrons with opposite spin at
small distances below one Bohr radius which is due to the missing
Pauli repulsion in this case. This trend increases with lowering
of the temperature due to increasing quantum effects.
Before analyzing the lowest temperature in Fig.~\ref{fig:Ee_pp_pdf} 
consider the electron-proton distributions. Multiplying these
functions by $r^{2}$ gives essentially the radial probability
density $W_{ep}(r)=r^2g_{ep}(r)$,
which is plotted in Fig.~\ref{fig:Ep_pdf}. At low temperatures
this function converges to the ground state probability density
of the hydrogen atom $W_{ep}(r)=r^2 (r)|\psi|^2_{1s}(r)$ influenced
by the surrounding plasma. Here,
lowering of the temperature leads towards the formation of a
shoulder around 1.4$a_{B}$ for $r_{s}=4$ and 1.2$a_{B}$ for
$r_{s}=6$ which is due to the formation of hydrogen atoms (which is 
confirmed by the corresponding quasi-bound electron
trajectories). At this temperature, the observed
most probable electron distance is slightly larger than one
$a_{B}$ as in the atom hydrogen ground state. Of course,
classical MD cannot yield quantization of the bound
electron motion, but it correctly reproduces (via averaging over the
trajectories) the statistical properties of the atoms.

At $62,500K$ and $r_{s}=6$ (right center part of Fig.~\ref{fig:Ee_pp_pdf}) the
simulations show a first weak signature of molecule formation --
see the maximum of the p-p distribution function around $r=2a_{B}$
and the maximum of the distribution function of electrons with
anti-parallel spins around $r=1.5a_{B}$.
Upon further lowering of the temperature by a factor of two (lower panel
of Fig.~\ref{fig:Ee_pp_pdf}) the p-p functions exhibit a clear peak very
close to $r=1.4a_{B}$ -- the theoretical p-p separation in $H_{2}$. 
At the same time, also the e-e functions have a clear
peak around $r=0.5a_{B}$ (the two electrons are concentrated between the 
protons). In contrast, in the case of parallel spins, no molecules are formed,
the most probable electron distance is around
$r=1.2a_{B}$.

\noindent
{\em MD results for dynamic quantities.} 
We now extend the analysis to the dynamic properties
of an hydrogen plasma in equilibrium which is based on the fluctuation-dissipation 
theorem. The time-dependent microscopic density of plasma
species $\alpha$ is defined as
$\;
\rho^{\alpha}(\mathbf{r},t)=\sum_{i=1}^{N_{\alpha}}\delta[\mathbf{r}-\mathbf{r}_{i}^{\alpha}(t)]\label{eq:microscopic
density}$, with the Fourier components
$\;
\rho^{\alpha}(\mathbf{k},t)=\sum_{i=1}^{N_{\alpha}}\exp[i\mathbf{k}\cdot\mathbf{r}_{i}^{\alpha}(t)]\label{eq:fourier
component of species}$, where $\vec{r}_i(t)$ denotes the trajectory of
particle ``i'' obtained in the simulation.
We now define the three
partial density-density time correlation functions (DDCF) between sorts
$\alpha$ and $\eta$ as
\begin{equation}
A^{\alpha\eta}(\mathbf{k},t)=(N_{\alpha}+N_{\eta})^{-1}\left\langle
\rho^{\alpha}(\mathbf{k},t)\rho^{\eta}(-\mathbf{k},0)\right\rangle,
\label{ddcf}
\end{equation}
where, due to isotropy, $\mathbf{k}=k$. Here $\langle \dots \rangle,$ 
denotes averaging along the trajectories by shifting the time interval and keeping the difference equal to $t$.  
Note also, that $A^{\alpha\eta}(k,t)=A^{\eta\alpha}(k,t)$ for all pairs $\alpha$ and $\eta$.
In addition to the spin-resolved electron functions we can also consider the
spin averaged correlation function
$A^{e}(k,t)= A^{\uparrow\uparrow}(k,t)+A^{\downarrow\uparrow}(k,t).$

\begin{figure}[htb]
\begin{minipage}[t]{.49\textwidth}
\hspace{-1cm} \vspace{0cm}
\includegraphics[%
  bb=0bp -80bp 612bp 792bp,
  scale=0.23,
  angle=-90]{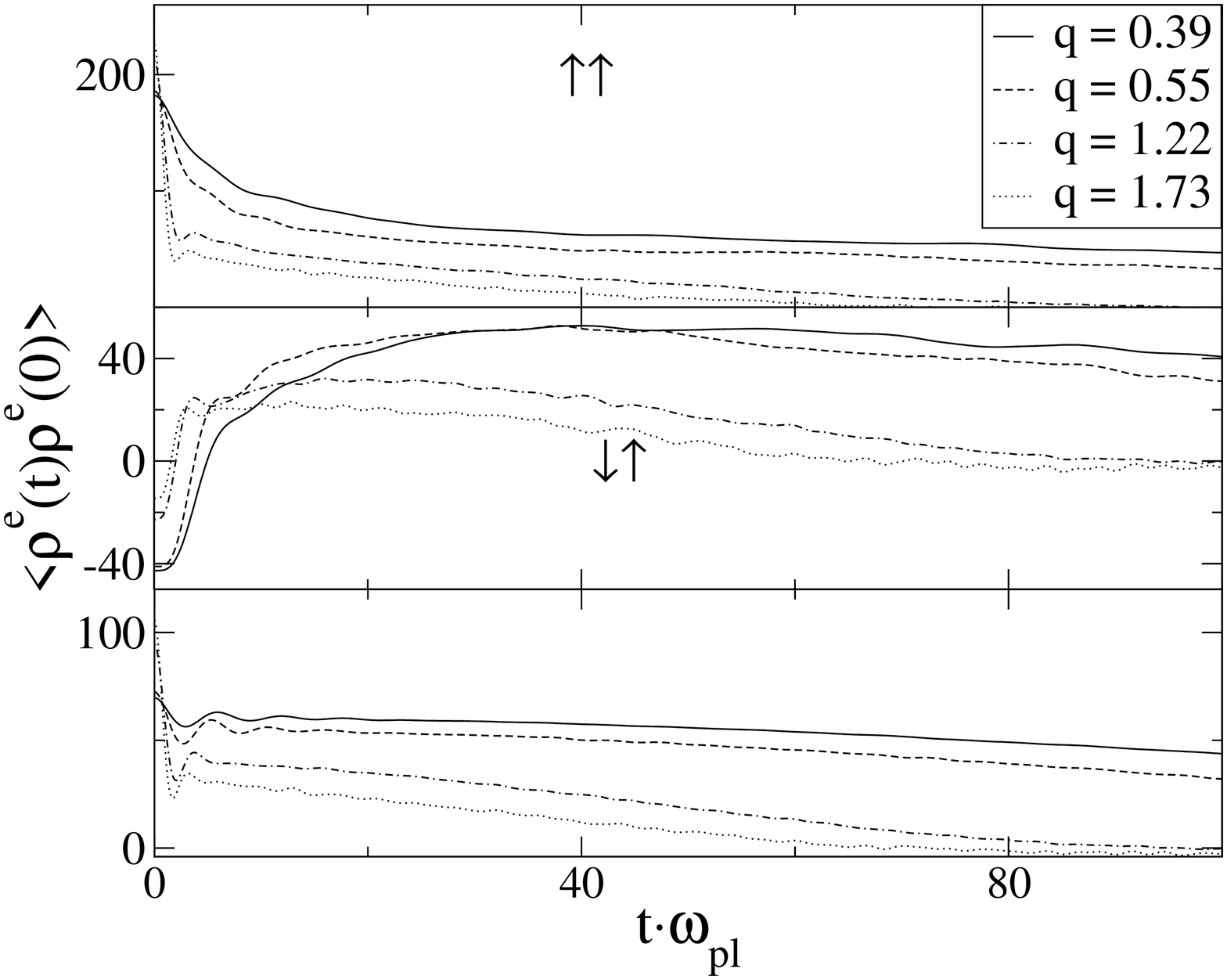}
\caption{Electron DDCF (\ref{ddcf}) multiplied by
$(N_{e}^{\uparrow}+N_{e}^{\downarrow})$ for $\Gamma=$ 1 and $\chi_e=1$ for four wave vectors.
Upper (center) figure: correlation functions for parallel (antiparallel) spins,
bottom figure: spin-averaged function.}
\label{fig:E_dcf}
\end{minipage}
\hfil
\begin{minipage}[t]{.43\textwidth}
\hspace{-1cm}
\includegraphics[%
  bb=0bp -50bp 612bp 792bp,
  scale=0.23,
  angle=-90]{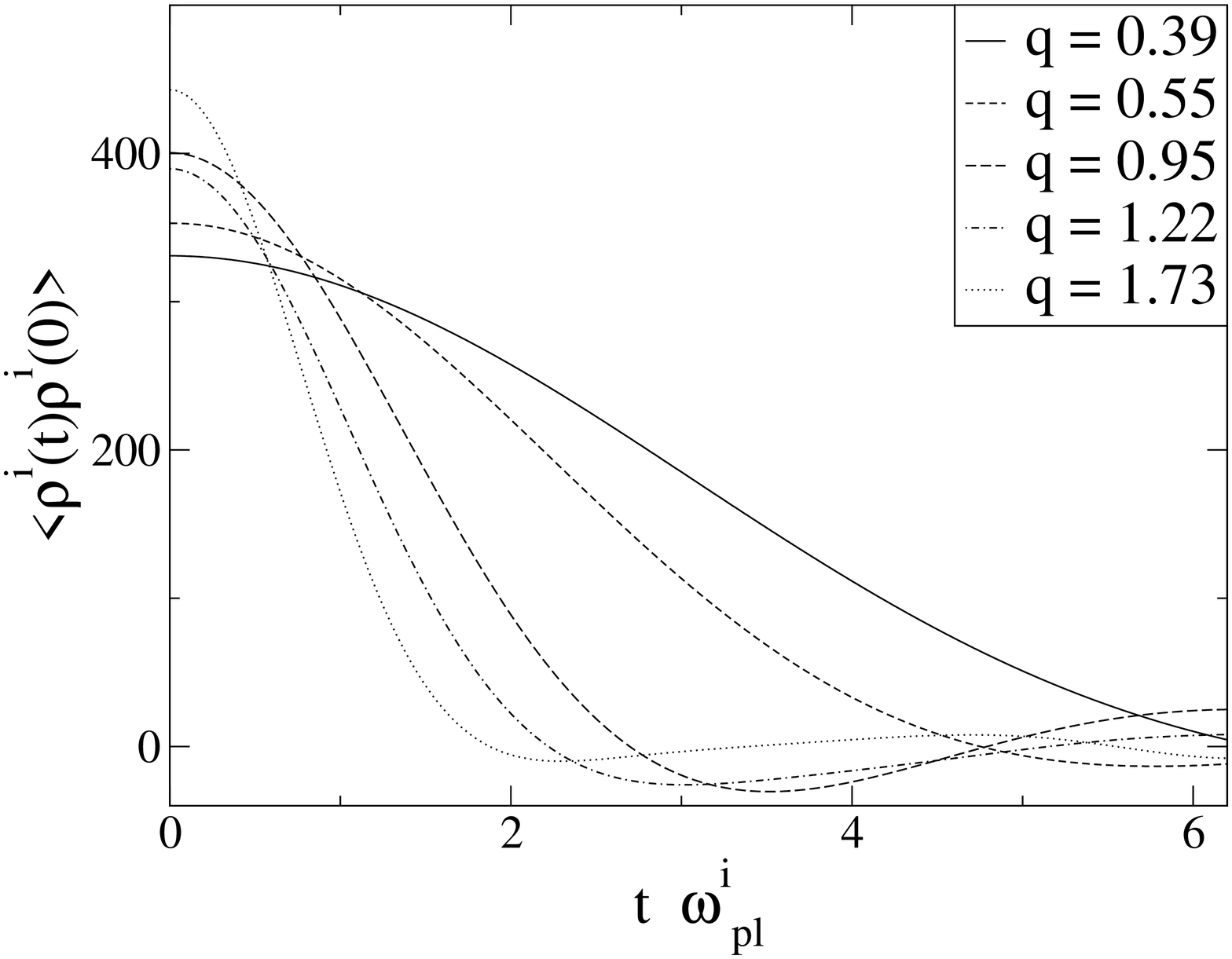}
\caption{Proton DDCF (\ref{ddcf}) for
$\Gamma=$ 1 and $\chi_e=1$ for five wave vectors (in units of $1/\bar{r}$).} 
\label{fig:P_dcf}
\end{minipage}
\end{figure}

We have performed a series of simulation runs
of equilibrium fluctuations in hydrogen plasmas with coupling
parameters $\Gamma$ and electron degeneracy parameters $\chi_e = \rho\Lambda_e^{3}$ with 
the elctron DeBroglie wavelength $\Lambda_e=\hbar/\sqrt{2\pi m_e k_BT}$ ranging from zero 
(classical system) to one (quantum or ``degenerate'' system).
The electron DDCF for $\Gamma=1$ and $\chi_e=1$ are plotted in
Fig.~\ref{fig:E_dcf} for four values of the dimensionless wavenumber,
$q=k\bar{\, r}$. The correlation functions ($\uparrow\uparrow$
and $\downarrow\uparrow$) have two characteristic behaviors -- a highly damped, 
high-frequency part and a weakly damped
low-frequency tail. The latter is related to the slow
ionic motion and the first one to oscillations with frequencies close to
the electron plasma frequency $\omega_{pl}$. On the other hand, the time scale of
the ion motion is determined by the ion plasma frequency
$\omega_{pl}^{i}=\sqrt{4\pi\rho_{i}Z_i^2 e^2/m_{i}}$, their ratio being about 43
(square root of the ion-to-electron mass ratio).
The slow proton oscillations are clearly seen in the proton DDCF,
shown in Fig.~\ref{fig:P_dcf}. To resolve the proton oscillations
the whole simulation (including the electron dynamics) has to extend over
several proton plasma periods $T_p=2\pi/\omega_{pl}^{i}$ thereby resolving the 
fast electronic motions as well, which sets the numerical limitation of the calculation.

\begin{figure}[htb]
\includegraphics[%
  scale=0.25, angle=-90]{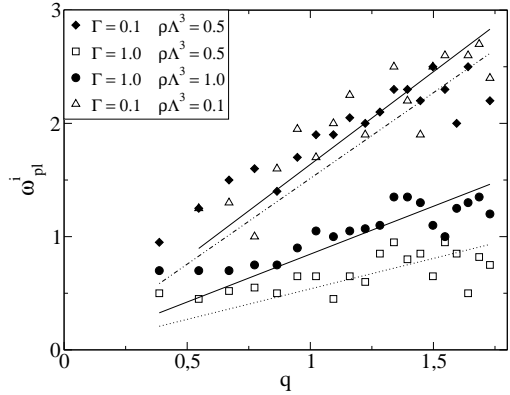}
\hspace{0.2cm}
\begin{minipage}[t]{.33\textwidth}
\caption{Ion-acoustic wave dispersion in a dense hydrogen plasma. Lines correspond to
weighted linear fits to the MD data (symbols). The scatter of the data is due to the limited 
particle number $N$ and simulation time and can be systematically reduced. Also, smaller q-values 
require larger $N$.}
\label{fig:P_dis}
\end{minipage}
\end{figure}

The temporal Fourier transform of the DDCF yields another very important quantity -- 
the {\em dynamic structure factor},
$S_{\alpha,\eta}(\omega,\, q)$, which allows one to analyze, e.g., the dispersion of the coupled
electron and proton oscillations. Fig.~\ref{fig:P_dis} shows dispersion results
for the collective proton oscillations (for the electron modes, see Refs.~\cite{vova01,golub2})
which follow from the peak positions of $S_{ii}(\omega,\, q)$. Fig.~\ref{fig:P_dis} shows
the peak frequency versus wave number, i.e. the dispersion of longitudinal ion-acoustic waves, 
$\omega(q)=v_{\rm MD}\cdot q$, where $v_{\rm MD}$ denotes our MD result for the phase velocity. 
This can be compared to the familiar analytical expression for an ideal two-temperature ($T_e \gg T_i$) plasma
$v_{s}=\sqrt{{Z_{i}k_{B}T_{e}}/{m_{i}}}$, where $v_s$ is the ion sound velocity. We observe deviations 
of about $10\%$, for weak degeneracy, $\chi_e < 0.5$, and about$10\%$, for large degeneracy, $\chi_e \ge 1$, 
which are due to nonideality (correlations) and quantum effects, directly included in our simulations. 
For further details on this method, see Refs.~\cite{golub2,bonitz-etal.05cpp,zwicknagel06}.

Thus the smiclassical MD is a powerful approach to correlated quantum plasmas. Thermodynamic and dynamic 
properties are accurately computed if accurate quantum pair potentials, such as the IKP, are being used.

\section{Quantum dynamics}\label{qd_s}

Now we discuss the method of Wigner trajectories in more detail.
As we have seen, the Wigner function $W$ [to avoid confusion, in this section we rename $F^W \rightarrow W$] in Eq.~(\ref{Fw_coord}) is the Fourier transform of 
the non-diagonal elements of the density matrix which, for a pure state, is 
$\rho\left(q +\frac{\nu}{2},q -\frac{\nu}{2} \right)=\psi( q+\frac
\nu 2,t) \psi ^{*}(q-\frac \nu 2,t) $, where the $N-$particle wave
functions satisfy the Schr\"odinger equation with an initial condition
\begin{eqnarray}
&&i\hbar \frac{\partial \psi }{\partial t}=\hat H \psi ;\quad \psi
\left( t_0\right) =\psi ^0\left( q\right) \label{s2},
\end{eqnarray}
which contains the hamiltonian (\ref{h}) [recall that $q$ is a vector of 
dimensionality $\{N\times d\}$].
By taking the time derivative of $W$ in Eq.~(\ref{Fw_coord}),
substituting it in the l.h.s of the Schr\"odinger equation instead
of $\partial \psi/\partial t$ and integrating by parts, we recover 
Eq.~(\ref{qwl}). For convenience of the further analysis, we separate the 
contribution of the classical force ${\vec F}(\vec q)=-\vec\nabla_q V(\vec q)$, 
compensating it in the function $\omega$ which now replaces $\tilde{\omega}$,
\begin{eqnarray}
\frac{\partial W}{\partial t}+ \frac{{\vec p}}{m} \cdot {\vec\nabla}_qW + {\vec
F}(\vec q) \cdot {\vec\nabla}_pW=
\int_{-\infty }^{\infty }ds\, W\left(
p-s,q,t\right) \, \omega \left( s,q,t\right), \qquad
\label{wig_eq}
\\
\omega \left( s,q,t\right) =\frac{2}{(\pi \hbar ^{2})^{N\, d}}\int
dq^{\prime }V\left( q-q^{\prime },t\right) \sin \left(
\frac{2sq^{\prime }}{\hbar } \right) + {\vec F}\left( q\right) \cdot {\vec\nabla}_s
\delta \left( s\right).  \qquad
\label{omega}
\end{eqnarray}
In the classical limit ($\hbar \rightarrow 0$), the r.h.s of
Eq.~(\ref{wig_eq}) vanishes and we obtain the classical Liouville
equation
\begin{equation}
\frac{\partial W}{\partial t}+\frac{{\vec p}}{m}\cdot{\vec\nabla}_qW+{\vec F}\left({\vec q}\right) \cdot{\vec\nabla}_pW = 0.\label{liuv}
\end{equation}

The solution of Eq.~(\ref{liuv}) is known and can be expressed by the Green function~\cite{Tatarsky1983}
\[
G(p,q,t;p_{0} ,q_{0},t_{0})=\delta [p-\overline{p}(t;t_{0} ,p_{0}
,q_{0} )]\delta [q-\overline{q} (t;t_{0},p_{0},q_{0} )],
\]
where $\overline{p}(\tau )$ and $\overline{q}(\tau )$ are the phase space
trajectories (of all particles), which are the solutions of Hamilton's equations together with the initial conditions at $\tau=t_0=0$,
\begin{eqnarray}
&&d\bar{q}/d\tau =\bar{p}(\tau )/m;\quad \bar{q}(0)=q_0, \nonumber \\
&&d\bar{p}/d\tau =\vec F(\bar{q}(\tau )); \quad \bar{p}(0)=p_0.
\label{gam}
\end{eqnarray}
Using the Green function, the time-dependent solution of the classical
Liouville equation takes the form
\begin{eqnarray}
W\left(p,q,t\right)= \int dp_{0} dq_{0} \, G(p,q,t;p_{0},q_{0},0)\, W_0(p_{0},q_{0}). \label{greencl}
\end{eqnarray}

With this result, it is now possible to construct a solution also for the quantum case. To this end we note 
that it is straightforward to convert Eq.~(\ref{wig_eq}) into an integral equation
\begin{eqnarray}
W\left(p,q,t\right)= \int dp_{0} dq_{0} \, G(p,q,t;p_{0},q_{0},0)\, W_0(p_{0},q_{0})+ \qquad\qquad\qquad
\label{green}
\\\nonumber 
 \int_{0}^{t}dt_{1}\int dp_{1}dq_{1} \,
G(p,q,t;p_{1},q_{1},t_{1})\int_{-\infty }^{\infty }ds_{1}\,
\omega (s_{1},q_{1},t_{1})\, W(p_{1}-s_{1},q_{1},t_{1}), 
\nonumber 
\end{eqnarray}
which is exact and can be solved efficiently by iterations~\cite{filmd0,ourbook}.
First, we can express the function $W(p_{1}-s_{1},q_{1},t_{1})$
written on the r.h.s. of Eq.~(\ref{green}), formally, using the
same Eq.~(\ref{green}) but written for another set of variables,
i.e. $\{p,q,t\}\rightarrow \{p_{1}-s_{1},q_{1},t_{1}\}$ and
$\{p_1,q_1,s_1,t_1\}\rightarrow \{p_{2},q_{2},s_2,t_{2}\}$. Second, substitution of the
obtained result again in Eq.~(\ref{green}) gives
\begin{eqnarray}
W\left( p,q,t \right)= W^{(0)}(p,q,t) +  W^{(1)}\left( p,q,t\right) +
\int_{0}^{t}dt_{1}\int d1\, G(p,q,t;1,t_{1})\times 
\nonumber \\
\int_{0}^{t_{1}}dt_{2}\int d 2\,
G(p_{1}-s_{1},q_{1},t_{1};2,t_{2})
\int_{-\infty}^{\infty }ds_{2}\, \omega (s_{2},q_{2},t_{2})\,
W(p_{2}-s_{2},q_{2} ,t_{2}),\nonumber\\
\label{green2}
\end{eqnarray}
where we have introduced the short notations $n\equiv q_n, p_n$, $d n\equiv dq_n dp_n$ and
\begin{eqnarray}
W^{(0)}(p,q,t) &=& \int d 0
G(p,q,t;0,0)W_0(0), 
\label{w20}\\
W^{(1)}\left( p,q,t\right) &=& \int_{0}^{t}dt_{1}
\int_{-\infty }^{\infty } d 1\,
G(p,q,t;1,t_{1})\int_{-\infty }^{\infty}ds_{1}\, \omega (s_{1},q_{1},t_{1}) 
\nonumber \\
&&\times\int d 0\, G(p_{1}-s_{1},q_{1},t_{1};0,0)W_0(0).
\label{w2}
\end{eqnarray}
$W^{(0)}(p,q,t)$ (as it follows from the Green function $G(p,q,t;p_{0},q_{0},0)$) describes the
propagation of the Wigner function along the classical characteristics, i.e., 
the solutions of Hamilton's equations~(\ref{gam}) in the time interval
$[0,t]$. It is worth mentioning, that this first term describes
both classical and quantum effects, due to the fact that the initial Wigner
function $W_0(p_{0},q_{0})$, in general, contains all powers of Planck's constant
$\hbar$ contained in the initial state wave functions. The second and third terms on the r.h.s. of
Eq.~(\ref{green2}) describe additional quantum corrections to the time
evolution of $W(p,q,t)$.

Let us consider the term $W^{(1)}(p,q,t)$ in more detail.
It was first proposed in Ref.~\cite{filmd0} and demonstrated
in Refs.~\cite{filmd1}$^-$\cite{filmd2} that
the multiple integral~(\ref{w2}) can be calculated stochastically
by Monte Carlo techniques.
For this we need to generate an ensemble of trajectories in
phase space. To each trajectory we ascribe a specific weight,
which gives its contribution to~(\ref{w2}). For example, let us consider a
trajectory which starts at point $\{p_{0},q_{0},\tau=0 \}$. This
trajectory acquires a weight equal to the value $W_0(p_{0},q_{0})$. Up to the time $\tau=t_1$ the
trajectory is defined by the Green function
$G(p_{1}-s_{1},q_{1},t_{1};p_{0},q_{0},0)$. At $\tau=t_1$, as it
follows from Eq.~(\ref{w2}), the weight of this trajectory must be
multiplied by the factor $\omega (s_{1},q_{1},t_{1})$, and
simultaneously a perturbation in momentum takes place:
$(p_{1}-s_{1}) \rightarrow p_1$. As a result the trajectory
becomes discontinuous in momentum space (but continuous in
the coordinate space). This is, obviously, a manifestation of the Heisenberg 
uncertainty of coordinates and momenta. Now, the trajectory consists of two parts -- two classical
trajectories which are the solutions of Eq.~(\ref{gam}), which are
separated, at $\tau=t_1$, by a ``momentum jump'' of magnitude $s_1$. 
What about the value $s_1$ of the jump and the time instance $t_1$? Both appear under 
integrals with a certain probability. To sample this probability adequately, a
statistical ensemble of trajectories should be generated, further the
instant of time $t_1$ must be chosen  randomly in the interval
$\left[0,t \right]$, and the momentum jump $s_1$ randomly in the interval
$\left(-\infty,+\infty \right)$. Finally, also different starting
points $\{p_0,q_0\}$ of trajectories at $\tau=0$ must be
considered (due to the integration $\int dp_{0}dq_{0}$).
Considering a sufficiently large number of trajectories of such type we can
accurately calculate $W^{(1)}\left( p,q,t\right)$ -- the first correction to the
classical evolution of the quantum distribution function
$W^{(0)}(p,q,t)$.

Let us now take into account the third term in
Eq.~(\ref{green2}). We now substitute, instead of
$W(p_{2}-s_{2},q_{2} ,t_{2})$, its integral representation,
using~(\ref{green}). As a result we get, for this term,
\begin{eqnarray}
W^{(2)}(p,q,t) = 
\int_{0}^{t}dt_{1} \int d 1 G(p,q,t;1,t_{1}) \int_{-\infty}^{\infty} ds_1 \, \omega(s_{1},q_{1},t_{1}) \times 
\nonumber \\
\int_{0}^{t_1} dt_{2} \int d 2\, G(p_{1}-s_{1},q_{1},t_{1};2,t_{2}) 
\int_{-\infty}^{\infty} ds_2\, \omega(s_{2},q_{2},t_{2}) \times  
\nonumber \\
\int d 0 G(p_{2}-s_{2},q_{2},t_{2};0,0)\, W_0(0). \qquad\qquad
\label{w3}
\end{eqnarray}
If we apply the stochastic interpretation of the integrals, as we
did above for $W^{(1)}\left( p,q,t\right)$, this term can be analogously
calculated using an ensemble of classical trajectories with {\em two} momentum jumps 
taking place at time instants $\tau=t_1$ and
$\tau=t_2$, and with a weight function multiplied by the
factors  $\omega (s_{1},q_{1},t_{1})$ and $\omega (s_{2},q_{2},t_{2})$, respectively.

\begin{figure}[htb]
\vspace{-1cm}
\centering\epsfig{file=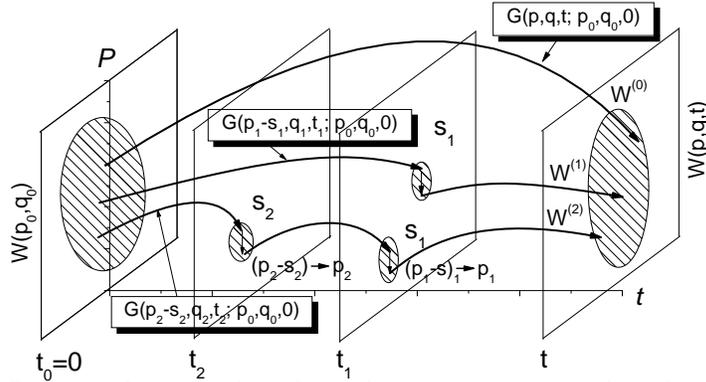,height=10cm,angle=-90}
\vspace{-1.5cm}\caption[]{Illustration of the iteration series. Three types of trajectories 
are shown: without (top curve), with one (middle) and two (lower) momentum jumps.}
\label{qmd_wc}
\end{figure}

Applying the above procedure several times, we can get the
higher order correction terms. As a result, $W(p,q,t)$ will be
expressed as an iteration series, with each term of the series
representing a contribution of trajectories of a definite
topological type -- with {\em one}, {\em two}, {\em three}, etc. momentum jumps. In
Fig.~\ref{qmd_wc} we show an example of trajectories
contributing to the terms $W^{(0)}, W^{(1)}$ and $W^{(2)}$.

As was noted in Section~\ref{wf_s} the Wigner function allows us to compute 
quantum-mechanical expectation values of an aribtrary one-particle operator
$\hat A$. Using the idea of the iteration series (\ref{green2}), we obtain an iteration series 
also for the expectation value,
\begin{eqnarray}
\langle \hat {A} \rangle (t) =\int dp dq \, A(p,q) W(p,q,t)=
\langle \hat {A} \rangle^{(0)}(t)+\langle \hat {A} \rangle^{(1)}(t)+\dots,
\label{a1}
\end{eqnarray}
where different terms correspond to different terms in the series for $W$.
The series (\ref{a1}) maybe computed much more efficiently than the one for $W$ since the result 
does not depend on coordinates and momenta anymore.

Certainly, in the iteration series it is possible to take into account
only a finite number of terms and contributions of a limited number of trajectories.
Interestingly, it is not necessary to compute the individual terms iteratively.
Instead, all (relevant) terms can be calculated simultaneously using the basic concepts 
of Monte Carlo methods, e.g.~\cite{Sobol}. An important task of the
Monte Carlo procedure will be to generate stochastically mostly
the trajectories which give the dominant contribution to the result, for details, see \cite{ourbook}.

\section{Time correlation functions in the canonical ensemble}

So far we have considered the dynamics of pure states where the density matrix
$\rho$ is defined by a single wavefunction $\psi$. However, at finite temperature
$\rho$ is, in general, defined by an incoherent superposition of wave
functions (mixed states). Here we consider a canonical ensemble as the
most common one. Time correlation functions $C_{FA}(t)=\left\langle
F(0)A(t)\right\rangle $ are among the most important quantities in
Statistical physics which describe transport properties, such as diffusion,
dielectric properties, chemical reaction rates, equilibrium and
non-equilibrium optical properties, etc. An example has already been considered in 
Sec.~\ref{MDsimulation} -- the density-density auto-correlation function (\ref{ddcf}).
Here, we use a more general expression for the quantum correlation function of two 
quantities $A$ and $F$ given by the operators $\hat F$ and $\hat A$. In the canonical
ensemble the averaging is performed by a trace with the canonical 
density operator ${\hat \rho}^{EQ}=Z^{-1}e^{-\beta \hat H}$, with $\beta=1/k_BT$, 
and the correlation function has the form~\cite{zubar}
\begin{eqnarray}
C_{FA}(t)=\frac{1}{Z}\text{Tr}(\hat F \, e^{i\hat H t_{\beta}^*}  \,
\hat A \,  e^{-i\hat H t_{\beta}}), \label{c_af}
\end{eqnarray}
where $\hat H$ is the Hamiltonian (\ref{h}), $t_{\beta}$ is a complex time argument (it ``absorbs'' ${\hat \rho}^{EQ}$),
 $t_{\beta}=t-i \beta/2$, $Z=\text{Tr}\rho^{EQ}$ is the partition function, 
and we use $\hbar=1$.

The time correlation function can now be computed by first, writing Eq.~(\ref{c_af}) in 
coordinate representation and then transforming to the Wigner picture using the 
Weyl representation of $\hat F$ and $\hat A$,
\begin{eqnarray}
&&C_{FA}(t)=\nonumber \\
&& Z^{-1}\int dq_1 dq_2 dq_3 dq_4\, \left\langle q_1 |\hat F| q_2
\right\rangle \, \left\langle q_2 | e^{i\hat H t_{\beta}^*} | q_3
\right\rangle \, \left\langle q_3 |\hat A| q_4 \right\rangle
\,\left\langle q_4 | e^{-i\hat H t_{\beta}} | q_1 \right\rangle \nonumber \\
&&= \int dp_1 dq_1 dp_2 dq_2\, F(p_1,q_1)\, A(p_2,q_2)\, W(p_1,q_1;
p_2,q_2; t; \beta), \label{c_fa}
\end{eqnarray}
where $W(p_1,q_1; p_2,q_2; t; \beta)$ is now a generalization of the Wigner function which is 
defined as double Fourier transformation of the product of two non-diagonal
matrix elements of the density operator
\begin{eqnarray}
&&W(p_1,q_1; p_2,q_2; t; \beta)=\frac{1}{Z (2 \pi)^{2Nd}} \int d\xi_1 d\xi_2 \, e^{i p_1\xi_1} \, e^{i p_2\xi_2}\times
 \nonumber \\
&& \hspace{0.5cm} \left\langle q_1 - \frac{\xi_1}{2} \left| e^{i\hat H t_{\beta}^*} \right| q_2 + 
\frac{\xi_2}{2} \right\rangle \,\left\langle q_2 -
\frac{\xi_2}{2} \left| e^{-i\hat H t_{\beta}} \right| q_1 +
\frac{\xi_1}{2} \right\rangle. 
\label{specf}
\end{eqnarray}
Calculating the partial time derivatives of the function $W$
using the usual technique for matrix elements, it can be shown
that the function $W$ satisfies a system of two Wigner-Liouville
equations~\cite{filmd1,filmd11}
\begin{eqnarray}
&&\frac{\partial W}{\partial t}+\frac{\vec p_1}{m}\cdot
\vec\nabla_{q_1} W +\vec F(\vec q_1) \cdot
\vec\nabla_{p_1} W=I_1, \nonumber \\
&&\frac{\partial W}{\partial t}+\frac{\vec p_2}{m} \cdot
\vec\nabla_{q_2} W +\vec F(\vec q_2) \cdot
\vec\nabla_{p_2} W=I_2, \label{wig_eq2}
\end{eqnarray}
where on the r.h.s. we have two ``collison'' integrals
\begin{eqnarray}
&&I_1=\int_{-\infty
}^{\infty }ds_1\, W\left( p_1-s_1,q_1; p_2,q_2; t; \beta \right)
\,\omega \left( s_1,q_1,t\right) \nonumber \\
&& I_2= \int_{-\infty
}^{\infty }ds_2\, W\left( p_1,q_1; p_2-s_2,q_2; t; \beta \right)
\,\omega \left( s_2,q_2,t\right),
\end{eqnarray}
and the function $\omega \left( s,q,t\right)$ is defined in the same
way as in the microcanonical ensemble, see Eq.~(\ref{omega}).

\subsection{Initial conditions for the Wigner-Liouville equation}

\begin{figure}
\centering\epsfig{file=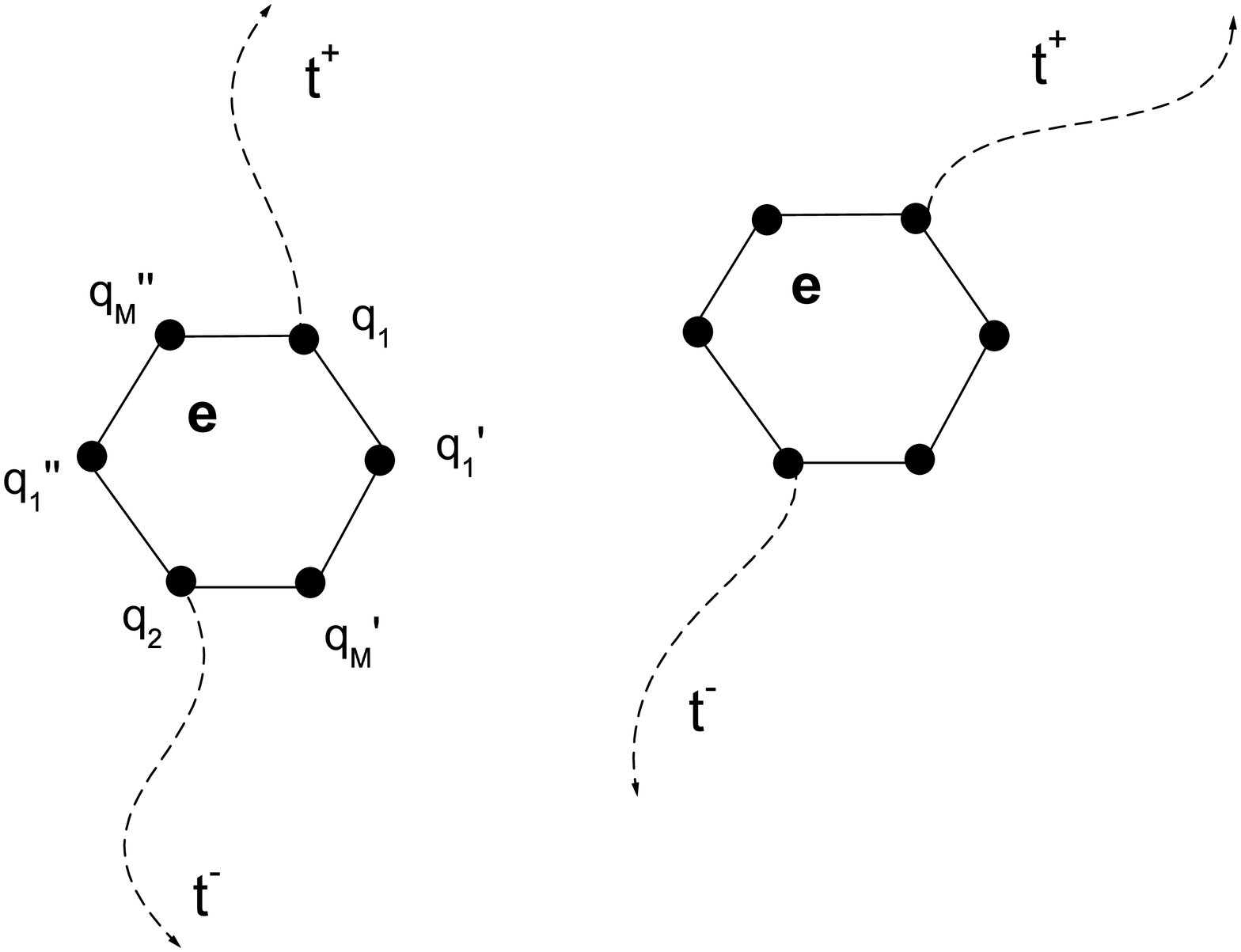,width=5.0cm,angle=-90}
\caption[]{Two closed loops illustrating the path
integral representation of two electrons in the density matricies in Eq.~(\ref{wapp}), 
see also Ref.~\cite{mchere}. Two special points $(p_1,q_1)$ and $(p_2,q_2)$. are
starting points for two dynamical trajectories propagating
forward and backward in time.}
\label{beads6}
\end{figure}

Using Eq.~(\ref{specf}) at $t=0$, we find that the initial value
of the Wigner function is given by the integral (again $1=q_1,p_1$)
\begin{eqnarray}
&&W_0(1; 2; 0; \beta)=\frac{1}{Z(2 \pi)^{2Nd}} \int d\xi_1
d\xi_2 \, e^{i p_1\xi_1} \, e^{i p_2\xi_2} \nonumber \\
&& \hspace{0.5cm} \left\langle q_1 - \frac{\xi_1}{2} \left| e^{-\beta\hat H/2} \right| q_2 + \frac{\xi_2}{2} \right\rangle \,\left\langle q_2 -
\frac{\xi_2}{2} \left| e^{-\beta\hat H/2} \right| q_1 +
\frac{\xi_1}{2} \right\rangle.
\end{eqnarray}
Let us now use the group property of the density operator $\hat \rho$
and the high temperature approximation for the matrix elements of
$\langle q' |\hat\rho|q\rangle$~\cite{mchere},
\begin{eqnarray}
&&\hspace{1.9cm} e^{-\beta \hat H}=\left[ e^{-\frac{\beta}{M}\hat H} \right]^{M} \nonumber \\
&&\left\langle q' \left| e^{- \frac{\beta}{2M} \hat H} \right| q'' \right\rangle \approx
\left\langle q' \left|e^{- \frac{\beta}{2M} \hat K} \right| q'' \right\rangle \,
\left\langle q' \left| e^{- \frac{\beta}{2M} \hat U} \right|q'' \right\rangle.
\label{fact}
\end{eqnarray}
As a result, we obtain
\begin{eqnarray}
&&\hspace{-0.5cm} W_0(1; 2; 0; \beta)
\approx \frac{1}{Z(2 \pi \hbar)^{2Nd}} \int dq_1' \ldots dq_M'\, dq_1''
\ldots dq_M'' \, e^{-\sum\limits_{m=2}^{M} K_{m} -\sum\limits_{m=1}^{M} U_m}\times \nonumber \\
&& \int d\xi_1 \,e^{i p_1\xi_1/\hbar} \, \left\langle q_M' \left| e^{-\frac{\beta}{2M} \hat K}
\right| q_1 + \frac{\xi_1}{2} \,\right\rangle
\left\langle q_1 - \frac{\xi_1}{2}  \left| e^{-\frac{\beta}{2M} \hat K}
\right|  q_1'' \right\rangle \times \nonumber \\
&& \int d\xi_2 \,e^{i p_2\xi_2/\hbar} \,\left\langle q_M'' \left|
e^{-\frac{\beta}{2M} \hat K}  \right|
q_2 + \frac{\xi_2}{2} \right\rangle \,
\left\langle q_2 - \frac{\xi_2}{2}  \left| e^{-\frac{\beta}{2M} \hat K}
\right|  q_1' \right\rangle, \label{wapp}
\end{eqnarray}
where
\begin{equation}
K_{m} = \frac{\pi}{\lambda_M^2} \left[ (q_{m}' -q_{m-1}')^2 + (q_{m}''
-q_{m-1}'')^2 \right], \quad U_m = \frac{\beta}{2M} \left[ U(q_m')+U(q_m'')
\right].
\nonumber
\end{equation}
Here we have assumed that $M \gg 1$, and 
$\lambda_{M}^2=\frac{2\pi \hbar^2 \beta} {m\,M}$
denotes the thermal de Broglie wave length
corresponding to the inverse temperature $\frac{\beta}{2M}$. A direct
calculation of the last two factors in Eq.~(\ref{wapp}) gives
\begin{eqnarray}
&&\int d\xi_1\,\,e^{i p_1\xi_1/\hbar}\, \left\langle q_M' \left| e^{-\frac{\beta}{2M}\hat K}
\right| q_1 + \frac{\xi_1}{2} \right\rangle \, \left\langle q_1 -
\frac{\xi_1}{2}  \left| e^{-\frac{\beta}{2M} \hat K} \right|  q_1''
\right\rangle \nonumber \\
&&= \left\langle q_M' \left| e^{-\frac{\beta}{2M} \hat K} \right| q \right\rangle \, \phi (p;
q_M', q_1) \, \left\langle q \left| e^{-\frac{\beta}{2M}\hat K/2M} \right| q_1 \right\rangle,
\end{eqnarray}
where
\begin{eqnarray}
\phi\left(p; q_M', q_1\right)=(2\lambda_M^2)^{Nd/2}\, \exp \left( - \frac{(
p\lambda_M/\hbar + i\pi(q'-q'')/\lambda_M)^2}{2 \pi} \right).
\end{eqnarray}
The final result for the Wigner function at $t=0$ can be written as
\begin{eqnarray}
&&W(1; 2; 0; \beta) \approx \nonumber \\
&&\int dq_1' \ldots dq_M'\, dq_1'' \ldots dq_M'' \, \Psi (1;2; q_1' \ldots q_M'; q_1'' \ldots q_M''; 0; \beta)\times \nonumber \\
&& \hspace{0.5cm} \phi (p_2; q_M',q_1'') \, \phi (p_1; q_M'',q_1'),
\label{wapp2}
\end{eqnarray}
where
\begin{eqnarray}
\Psi (p_1,q_1; p_2,q_2; q_1' \ldots q_M'; q_1'' \ldots q_M'';
\beta)=Z^{-1}e^{-\sum\limits_{m=1}^{M+1} K_{m} -\sum\limits_{m=1}^{M} U_m}.
\nonumber
\end{eqnarray}
Here we have introduced the notation $\{q_0'\equiv q_1;
q_0''\equiv q_2\}$ and $\{q_{M+1}'\equiv q_2; q_{M+1}''\equiv q_1\}$.
The Fig.~\ref{beads6} is an illustration of the simulation idea. Two closed loops
with the set of points show the path integral representation of the density
matricies in Eq.~(\ref{wapp}). The left chain of points, i.e.\\
$\{q_1,q_1',\ldots,q_M',q_2,q_1'',\ldots,q_M''\}$ characterizes the ``path''
of a single quantum particle.
The chain has two special points
$(p_1,q_1)$ and $(p_2,q_2)$. As it follows from
Eqs.~(\ref{specf}),(\ref{wig_eq2}), these
points are the original points for the Wigner function (the additional arguments
arise from the path integral representation).
As we show in the next section, we can consider these points as
starting points for two dynamical trajectories propagating
forward and backward in time, i.e. $t\rightarrow t^+$ and $t\rightarrow t^-$.
The Hamilton equations for the trajectories are defined in the next section.

\subsection{Integral equations}

The solution scheme follows the one explained before.
The only difference is that we now have to propagate two trajectories instead of
one, %. The integral equation for the Wigner function are obtained from
%Eqs.~(\ref{wig_eq2}). Let us define a
%pair of trajectories propagating forward and backward in time
\begin{eqnarray}
&&\frac{d\bar{q}_1}{d\tau} =\frac{\bar{p}_1(\tau )}{2m};\hspace{1.55cm} \bar{q}_1(0)=q_1^{0} \nonumber \\
&&\frac{d\bar{p}_1}{d\tau} =\frac{1}{2} \, \vec F[\bar{q}_1(\tau )]; \hspace{0.7cm} \bar{p}_1(0)=p_1^{0} \nonumber \\
&&\frac{d\bar{q}_2}{d\tau} =-\frac{\bar{p}_2(\tau )}{2m};\hspace{1.25cm} \bar{q}_2(0)=q_2^{0} \nonumber \\
&&\frac{d\bar{p}_2}{d\tau} =-\frac{1}{2} \, \vec F[\bar{q}_2(\tau )];
\quad \bar{p}_2(0)=p_2^{0}. \label{gam2}
\end{eqnarray}
the first (second) propogating forward (backward)
Let us substitute expressions for $\vec F[\bar{q}_1(\tau )],
\bar{p}_1(\tau ), \vec F[\bar{q}_2(\tau )]$ and $\bar{p}_2(\tau )$
from~(\ref{gam2}) into Eqs.~(\ref{wig_eq2}) and subtract the
second equation from the first. As a result, in the l.h.s. we will
obtain a full differential of the Wigner function. After multiplication by
the factor $\{1/2\}$ and integration over time, the integral
equation for the Wigner function takes the form
\begin{eqnarray}
&&W\left(p_1,q_1; p_2,q_2; t; \beta \right)= \nonumber \\
& = &\int dp_1^{0} dq_1^{0} \, dp_2^{0} dq_2^{0} \; G(p_1,q_1,
p_2,q_2, t; p_1^{0},q_1^{0}, p_2^{0},q_2^{0}, 0) \;
 W(p_1^{0},q_1^{0}; p_2^{0}, q_2^{0}; 0; \beta)+ \nonumber \\
&& \int_{0}^{t} d\tau \int dp_{1}^{1}dq_{1}^{1}\,
dp_{2}^{1}dq_{2}^{1} \;
G(p_1,q_1, p_2,q_2, t; p_1^{1},q_1^{1}, p_2^{1},q_2^{1}, \tau)\times \nonumber \\
&&\int_{-\infty }^{\infty }ds \, d\eta \;
\vartheta(s,q_{1}^{1}; \eta, q_{2}^{1};\tau) \;
W(p_1^{1}-s,q_1^{1}; p_2^{1}-\eta,q_2^{1};\tau; \beta),
\label{green22}
\end{eqnarray}
where $\vartheta(s,q_{1}^{1}; \eta, q_{2}^{1};\tau)=\frac{1}{2}
\{\omega(s,q_{1}^{1})\delta (\eta)- \omega(\eta,q_{2}^{1})\delta
(s) \}$. The dynamical Green function $G$ is defined as
\begin{eqnarray}
&&\hspace{-1.4cm} G(p_1,q_1, p_2,q_2, t; p_1^{0},q_1^{0}, p_2^{0},q_2^{0}, 0) = \delta \{p_1 - \bar{p_1}(\tau; p_1^{0},q_1^{0},0)\}
\times \nonumber \\
&& \hspace{-1.4cm}\delta \{q_1 - \bar{q_1}(\tau; p_1^{0},q_1^{0},0)\} \cdot
\delta \{ p_2 - \bar{p_2}(\tau; p_2^{0},q_2^{0},0) \} \cdot  \delta
\{q_2 - \bar{q_2}(\tau; p_2^{0},q_2^{0},0)\}
\end{eqnarray}
Let us denote the first term on the r.h.s. of
Eq.~(\ref{green22}) as $W^{(0)}(p_1,q_1; p_2,q_2; t; \beta )$.
This term represents the Wigner function of the initial state
propagating along classical trajectories (characteristics),
solutions of Eqs.~(\ref{gam2}). Using the same approach which
was applied in the microcanonical ensemble, we obtain
expressions for
\[
W^{(1)}(p_1,q_1; p_2,q_2; t; \beta ), W^{(2)}(p_1,q_1; p_2,q_2; t;
\beta ), \ldots
\]
 and represent $W\left(p_1,q_1;p_2,q_2; t;
\beta \right)$ as an iteration series. In
this case, we can calculate this also with an
ensemble of trajectories using the Quantum-Dynamics-Monte-Carlo approach
which was described in~\cite{fil_koch}.
As a result the expression for the time
correlation function~(\ref{c_fa}) can be rewritten as
\begin{eqnarray}
&&C_{FA}(t)= \int dp_1 dq_1 \, dp_2 dq_2 \, F(p_1,q_1) \, A(p_2,q_2) \,
W(p_1,q_1; p_2,q_2; t; \beta)= \nonumber \\
&&\hspace{1.3cm} = \left( \phi(P) | W^{(0)}(P;\beta) \right) + \sum_{i=1}^{\infty }
\left( \phi(P) | W^{(i)}(P;\beta) \right), \label{c_fa2}
\end{eqnarray}
where $( \ldots|\ldots )$ denotes the integral in the (now $2N$-particle) 
phase space $\{p_1,q_1,p_2,q_2 \}$, and $\phi(P)= F(p_1,q_1)\, A(p_2,q_2)$.

An illustrative example for the calculations of the time correlation functions
$C_{FA}$ is the  momentum-momentum autocorrelation functions $C_{PP}(t)$ 
for a one-dimensional system of interacting electrons in an array of fixed random
scatterers at finite temperature~\cite{fil_koch}.
This system is of high interest because at zero temperature it shows
Anderson localization if
electron-electron (e-e) interaction is neglected. It has been a long
standing of the question what the effect of e-e interaction on
localization will be. The present method~\cite{fil_koch} is well
suited to answer this question, and it could be shown that  Coulomb e-e interaction
enhances the mobility of localized electrons~\cite{fil_koch,ourbook}.

\section{Discussion}
We have presented a general idea how to extend the powerful method of molecular dynamics to quantum systems. 
First, we discussed ``semiclassical MD'', i.e., classical MD with accurate quantum pair potentials.
This method is very efficient and allows to compute thermodynamic properties of partially ionized 
plasmas for temperatures above the molecule binding energy (i.e. as long as three and four particle correlations 
can be neglected). Further, frequency dependent quantities (e.g., plasmon spectrum) are computed correctly for 
$\omega < \omega_{pl}$. Further progress is possible if more general quantum potentials are derived. 

In the second part, we considered ways to rigorously solve the quantum Wigner-Liouville equation for the 
N-particle Wigner function. Results were derived for both, a pure quantum state and a mixed state (canonical ensemble).
Although this method is by now well formulated, it is still very CPU time costly, so that practical applications 
are only starting to emerge. Yet, we expect that, due to its first principle character, Wigner function QMD will 
become increasingly important for a large variety of complex many-body problems.

This work is supported by the Deutsche Forschungsgmeinschaft via SFB-TR 24 and in part by Award
No. Y2-P-11-02 of the U.S.
Civilian Research and Development Foundation for the Independent
States of the Former Soviet Union (CRDF) and of Ministry of Education and Science of
Russian Federation, and RF President Grant NS-3683.2006.2 for governmental support of
leading scientific schools.

\end{document}